\documentclass[]{emulateapj-rtx4}

\usepackage{graphicx,times}
\shorttitle{Testing diffusive core overshooting mixing on HY Vir}
\shortauthors{Q.S. Zhang}

\begin{document}

\title{Testing the core overshooting mixing described by the turbulent convection model on the eclipsing binary star HY Vir }
\author{Q.S. Zhang\altaffilmark{1,2,3}}
\email{zqs@ynao.ac.cn(QSZ)}

\altaffiltext{1}{National Astronomical Observatories/Yunnan Observatory, Chinese Academy of Sciences, P.O. Box 110, Kunming 650011, China.}
\altaffiltext{2}{Key Laboratory for the Structure and Evolution of Celestial Objects, Chinese Academy of Sciences, Kunming, 650011, China.}
\altaffiltext{3}{Graduate School of Chinese Academy of Sciences, Beijing 100039, China.}

\begin{abstract}
Helioseismic investigation has suggested to apply the turbulent convection models (TCMs) to the convective overshooting. Using the turbulent velocity in the overshooting region determined by the TCM, one can deal with the overshooting mixing as a diffusion process, which leads to incomplete mixing. It has been found that this treatment can improve the solar sound speed and the Li depletion in open clusters. In order to investigate whether the TCM can be applied to the overshooting mixing outside the stellar convective core, new observations of the eclipsing binary star HY Vir are adopted to calibrate the overshooting mixing parameter. The main conclusions are as follows: (i) the TCM parameters and the overshooting mixing parameter in the solar case are also suitable for the eclipsing binary system HY Vir; (ii) the incomplete mixing results in a continuous profile of the hydrogen abundance; (iii) the e-folding length of the region, in which the hydrogen abundance changes due to the overshooting mixing, increases during the stellar evolution.
\end{abstract}


\section{Introduction}

The convective overshooting in the stellar interior is caused by turbulent flows moving across the standard convective boundary, which is defined by the Schwarzchild criterion (i.e., $\nabla_R\geq\nabla_{ad}$), into the dynamically stable region. This process leads to chemical mixing in the overshooting region. Based on the framework of non-local mixing length theories, the overshooting region is considered as a fully mixed region with a length $l_{OV}=\alpha_{OV}H_P$, where $H_P$ is the local pressure scale height and $\alpha_{OV}$ is a parameter. But the helioseismic investigation has pointed out that only turbulent convection models (TCMs) can properly describe the convective overshooting \citep{chr11}. Therefore, it is a tendency to apply TCMs to the convective overshooting in the stellar interior. TCMs are non-local convection theories based on the hydrodynamic equations and modeling assumptions \citep{xio81,xio85,xio89,can97,can98,xio01,li01,den06,li07,can11,li12}. Using the turbulent velocity determined by TCMs, one can deal with the overshooting mixing as a diffusion process which results in incomplete mixing in the overshooting region.
\citet{zha12a} have applied the TCM \citep{li07} and the incomplete overshooting mixing to the downward overshooting region of the solar convective envelope and found that the sound speed profile is significant improved. Applying the same method to the stellar model also shows improvements on the Li depletion in open cluster, thus the TCM has been considered to be generally applicable to the downward overshooting region of the convective envelope for low-mass stars \citep{zha12}.

It is interesting whether the incomplete overshooting mixing based on the TCM is suitable for the convective core overshooting.
The convective core overshooting refreshes hydrogen in the nuclear burning core, deeply affecting the evolutions of massive and intermediate-mass stars. The evolution track of a star with a convective core in the Hertzsprung-Russel diagram is sensitive to the efficiency of the core overshooting mixing. Accordingly, it is a direct and efficient way of studying the core overshooting mixing by calibrating the radius and effective temperature of the stellar model to fit the observations. Double lined detached eclipsing binary systems in the main sequence phase are the best candidates to do that, because the mass, radius, and temperature of each component of a binary system can be obtained by observations. \citet{ri00} and \citet{cl07} have studied the core overshooting based on a set of double lined eclipsing binary systems and found that the fully mixed overshooting region with a length $0.20\sim0.25H_P$ is the overall best. Recently, \citet{lf11} have observed the double lined detached eclipsing binary system HY Vir and acquired very accurate masses and radii of the two components from the analysis of new light curves and radial velocity curves. The mass of each component of HY Vir is \citep{lf11}: $M_1=1.838\pm0.009M_{\odot}$ and $M_2=1.404\pm0.006M_{\odot}$ for the primary and the secondary, respectively. These results indicate that both the primary and the secondary have the convective core. The accurate data make HY Vir a good candidate to study the core overshooting mixing described by the TCM.

In this paper, I investigate the suitable parameter of the overshooting mixing via calibrating the effective temperatures and the radii of the stellar models of HY Vir to fit the observations, and study the effects of the application of the TCM \citep{li07}. The method of applying the TCM in the stellar models is described in Section 2. The calibrated results for the system of HY Vir are shown in Section 3. The properties of the stellar models based on the TCM are discussed in Section 4. The conclusions and discussions are presented in Section 5.

\section{The method of calculation }

\subsection{The overshooting mixing}

In the traditional way, the overshooting is assumed to result in the completely mixed region with the length being proportional to the local pressure scale height. This description of the overshooting is based on the framework of the 'ballistic' overshoot models, which are non-local mixing length theories(e.g., \citet{ss73,ma75,br81,zah91}). Those non-local mixing length theories result in a steep temperature gradient profile at the boundary of the overshooting region (e.g., \citet{ss73,zah91}). However, the recent helioseismic investigation has found that the required temperature gradient in the overshooting region is not in agreement with the framework of the 'ballistic' overshoot models, but suggests that only TCMs are capable of providing required temperature gradient profile \citep{chr11}.

According to the hydrodynamic equations, the chemical mixing is determined by the turbulent velocity - turbulent concentration correlation \citep{xio86,can99,can11}. Unfortunately, there is a lack of the knowledge for such a correlation in the astrophysical case. Besides the fully mixing, the incomplete mixing described by the diffusion is at present popular on dealing with the overshooting mixing (e.g., \citet{den96a,den96b,fre96,ven98,her00,pax11}). In the diffusion description, the turbulent velocity in the overshooting region is required to calculate the diffusion coefficient. The turbulent velocity is usually set as an exponentially decreasing function in the overshooting region (e.g., \citet{fre96,ven98}), and can be calculated by defining the exponential index and the initial value (i.e., the turbulent velocity at the convective boundary). An option to determine the initial value is extrapolation from the mixing length theory (MLT) (e.g., \citet{fre96,ven98}). However, the MLT is a local theory, while the non-local effect dominates near the convective boundary. Another option is to obtain the turbulent velocity by solving the TCM, which is non-local turbulent convection theory.

The diffusion coefficient is proportional to the characteristic length of mixing and the characteristic velocity:
$D = flv$, where $l$ is the characteristic length, $v$ is the characteristic turbulent velocity, and $f$ is a diffusion parameter. $v$ is assumed as $\sqrt{k}$, where $k$ is the turbulent kinetic energy. The diffusion parameter multiplying the characteristic length is assumed to be proportional to $H_P$, i.e., $fl= C_X H_P$. Therefore, the diffusion coefficient of the overshooting mixing is assumed as follow:
\begin{eqnarray}
D_{OV}=C_X H_P \sqrt{k}
\end{eqnarray}%

The characteristic length of mixing $l$ is between the Kolmogorov scale and the largest eddy scale. In order to cause incomplete mixing in the overshooting region in a time scale comparable with the evolutionary time scale, the characteristic length of mixing should be much smaller than the largest eddy scale \citep{den96a}. By analogy with Eq.(27) \& (29) in \citet{den96a}, $C_X$ in Eq.(1) should be on the magnitude order of $10^{-10}$, which is suitable for the downward overshooting region of the convective envelope in low-mass star \citep{zha12a,zha12}.

The diffusion equation of the hydrogen abundance in the stellar interior is as follow:
\begin{eqnarray}
\frac{\partial_{Mix} X}{\partial t}=\frac{\partial}{\partial M_r}[(4\pi \rho r^2)^2 D \frac{\partial X}{\partial M_r}]
\end{eqnarray}%
where $(\partial_{Mix} X /{\partial t})$ means the time derivative of hydrogen abundance caused by only convective and overshooting mixing, and $D$ is the diffusion coefficient due to convection or overshooting:
\begin{eqnarray}
D = \left\{ {\begin{array}{*{20}{c}}
   {{D_{OV}};({\nabla _R} < {\nabla _{ad}})}  \\
   {{D_{CZ}};({\nabla _R} \ge {\nabla _{ad}})}  \\
\end{array}} \right.
\end{eqnarray}%
where $D_{CZ}$ is diffusion coefficient in the convection zone. In order to ensure fully mixing in the convection zone, large $D_{CZ}$ is required so that $D_{CZ}=H_P \sqrt{k}$ is adopted in the calculations. It is not difficult to solve Eq.(2) by using the tridiagonal matrix algorithm.

\subsection{The stellar evolution code}

The stellar evolution code, which is originally developed by
\citet{paz69} and Kozlowski and updated by Sienkiewicz, is used to model the components of HY Vir.
The OPAL equation of state \citep{rog96}, the OPAL opacity tables for high temperatures \citep{igl96},
and the Alexander's opacity tables for low temperatures \citep{ale94} are used. The composition mixture is assumed to be the same as the solar mixture \citep{gs98}.

In this paper, I solve the complete structure equations comprising the stellar structure equations and the TCM \citep{li07} equations. In the stellar structure equations, the temperature gradient $\nabla$ is calculated as follow:
\begin{equation}
\nabla=\nabla_{R}-\frac{H_P}{T}\frac{\rho c_{P}}{\lambda}\overline{u_{r}^{\prime }T^{\prime }}
\end{equation}%
where $\overline{u_{r}^{\prime }T^{\prime }}$ is the turbulent heat flux determined by the TCM.

The evolution code has been modified to take into account the overshooting mixing. In the original code, the evolution of chemical composition in the stellar interior and the stellar structure are calculated separately. On each time step, the evolution of chemical composition (nuclear burning, complete mixing in the convection zone) is calculated before solving the stellar structure equations. In the present work, the procedure of solving Eq.(2) is performed at first. After the evolution of chemical composition (nuclear burning and the convective and overshooting mixing) during the time step being accomplished, the complete structure equations are solved, and the stellar structure variables and the turbulent variables are updated. The scheme of solving the complete structure equations is as follows:

(1) Solve the TCM equations based on the current stellar structure variables (e.g., $\rho,T,r,L$). Calculate the temperature gradient $\nabla$ at all mesh points according to Eq.(4).

(2) Solve the localized TCM in which the non-local terms (i.e., the diffusion terms) are ignored. Calculate the local temperature gradient $\nabla_L$ at all mesh points. Some discussions about the localized TCM can be found in Appendix A in \citet{zha12b}.

(3) Calculate the ratio $\eta =\nabla / \nabla_L$ at all mesh points. Use $\eta$ to get $\eta'$ at all mesh points based on the relaxation technic, i.e., $\eta'=\eta'_{pre}+\xi (\eta-\eta_{pre})$, with a proper relaxation parameter $\xi$, where $\eta'_{pre}$ and $\eta_{pre}$ are the previous value of $\eta'$ and $\eta$.

(4) Solve the stellar structure equations in which the temperature gradient is calculated as $\nabla = \eta' \nabla_L$, and update stellar structure variables (e.g., $\rho,T,r,L$). It should be noticed in this process that $\nabla_L$ is updated when the stellar structure variables are updated in the Newtonian iterations.

(5) Check differences of $|\eta-\eta'|$ and $|\eta-\eta_{pre}|$. The calculations are thought to converge if $|\eta-\eta'|$ and $|\eta-\eta_{pre}|$ are less than required accuracy at all mesh points, otherwise return to step (1).

It is not difficult to understand that the turbulent variables and the stellar structure variables are consistent with each other when $\eta_{pre}=\eta=\eta'$, thus the scheme is reasonable. The scheme is an improved version of the previous one (e.g., \citet{zha09}). It is found that this scheme is more stable and more efficient than the previous one.

There are seven parameters in the TCM \citep{li07}: $\alpha,C_t,C_e$ are dissipation parameters of turbulent kinetic energy $k$, turbulent heat flux $\overline{u_{r}^{\prime }T^{\prime }}$, and turbulent temperature fluctuation $\overline{T^{\prime }T^{\prime }}$, respectively; $C_s,C_{t1},C_{e1}$ are diffusion parameters of $k$, $\overline{u_{r}^{\prime }T^{\prime }}$ and $\overline{T^{\prime }T^{\prime }}$, respectively; $C_k$ is the parameter of 'return to isotropy' term.

\begin{table}
\begin{center}
\caption{The turbulent parameters adopted in this paper. }
\begin{tabular}{crrrrrr}
  \hline
   $C_t$ & $C_e$& $C_s$& $C_{t1}$& $C_{e1}$& $C_k$& $\alpha$\\
  \hline
  $7.5$   &$0.2$   &$0.08$   &$0$   &$0$   &$2.5$& $0.85$\\
  \hline
\end{tabular}
\end{center}
\end{table}

The turbulent parameters adopted in this paper are listed in Table 1. $C_s,C_k,C_t,C_e$ are the same in the case of the solar model \citep{zha12a}. The diffusion parameters $C_{t1}$ and $C_{e1}$ are set to zero, because the calculation doesn't converge when the diffusions of $\overline{u_{r}^{\prime }T^{\prime }}$ and $\overline{T^{\prime }T^{\prime }}$ are taken into account. The turbulent dissipation parameter $\alpha=0.85$ is adopted. This value is due to the solar calibration based on the input physics and other TCM parameters.

The age and chemical composition of each component of HY Vir are assumed to be the same. Interactions between the primary and the secondary are ignored. The stellar models evolve from the ZAMS to the age when the radius of the primary stellar model fits the observation. The time step is no more than 0.5\% of the age. All stellar models comprise more than 2000 mesh points.

\section{Calibrations of stellar models of HY Vir}

The main aim of this paper is to test whether the solar turbulent parameters (i.e., the TCM parameters in Table 1 and the overshooting mixing parameter $C_X=10^{-10}$) is suitable for the core overshooting region and reproduce the observations of HY Vir. Mathematically, only one turbulent parameter can be derived by calibrating the effective temperatures $T_{eff}$ and radii $R$ of stellar models for HY Vir. I choose the most sensitive parameter, i.e., $C_X$, to be adjustable. The strength of the overshooting mixing is determined by the diffusion coefficient, which is proportional to $C_X$ and $\sqrt{k}$. Besides $C_X$, the TCM parameters indirectly affect the diffusion coefficient because of their effects on $k$. However, in an acceptable range of the TCM parameters, $k$ doesn't change too much.

The adjustable parameters in modeling the components of HY Vir are as follows: the overshooting mixing parameter $C_X$, the initial hydrogen abundance $X$, and the metal abundance $Z$. The observations make it possible to fix the adjustable parameters ($C_X$, $X$, $Z$). The effective temperatures and the radii of the stars (i.e., $T_1$ and $R_1$ for the primary,  $T_2$ and $R_2$ for the secondary) are the functions of the adjustable parameters ($C_X$, $X$, $Z$) and the age $t$. There are four free independent variables and the same number of dependent variables. Because the effective temperatures and the radii should be consistent with the observations, the adjustable parameters and the age can be solved in the sense of math.

The scheme of solving the adjustable parameters is as follows:
(1) Use the trial parameters set ($C_X$, $X$, $Z$) to model the primary of HY Vir.
(2) Obtain the effective temperature $T_1$ and the age $t$ when the radius of the primary $R_1$ evolves to the observational value.
(3) Use the same parameters set ($C_X$, $X$, $Z$) to model the secondary of HY Vir. The stellar model evolves from the ZAMS to the age $t$. Obtain the effective temperature and the radius, i.e., $T_2$ and $R_2$, of the stellar model of the secondary at age $t$.
(4) Compare $lgT_1$, $lgT_2$ and $R_2$ with the observations shown in Table 2. Use the Newtonian iteration method to revise the parameters ($C_X$, $X$, $Z$), then go to step (1) until $lgT_1$, $lgT_2$, and $R_2$ are consistent with the observations within the accuracy of $0.001$.
$R_1$ is calibrated to fix the age $t$, because the radius of the primary is the most sensitive variable, i.e., $| dR_1 / dt | > Max(| dR_2 / dt |, | dlgT_1 / dt |, | dlgT_2 / dt |)$. Therefore, to derive $t$ by calibrating $R_1$ is the most accurate way comparing with calibrating $R_2$, $T_1$, and $T_2$.

\begin{table}
\begin{center}
\caption{The observations of HY Vir and the standard errors \citep{lf11}. }
\begin{tabular}{crrrr}
  \hline
  Parameter & Primary & Secondary\\
  \hline
  $M/M_{\odot}$   & $1.838\pm0.009$ & $1.404\pm0.006$  \\
  $lg(T_{eff})$ & $3.836\pm0.008$ & $3.816\pm0.008$  \\
  $R/R_{\odot}$   & $2.806\pm0.008$ & $1.519\pm0.008$  \\
  $lg(L/L_{\odot})$   & $1.20\pm0.04$ & $0.58\pm0.04$  \\
  \hline
\end{tabular}
\end{center}
\end{table}

Using the above scheme, the adjustable parameters and their standard errors are worked out as follows:
\begin{eqnarray}
C_X=(0.8\pm 0.5) \times 10^{-10}
\end{eqnarray}%
\begin{eqnarray}
X=0.67\pm 0.03
\end{eqnarray}%
\begin{eqnarray}
Z=0.031 \pm 0.007
\end{eqnarray}%

For the calibrated models, the age of the system is $t=1.4Gyr$. Those are consistent with Lacy \& Fekel's(2011) results, which show $Z=0.027$, $\Delta Y /\Delta Z =2.0$ ($Y$ is the initial helium abundance) and $t=(1.35 \pm 0.1)Gyr$.

The standard errors $\sigma$ of the parameters ($C_X$, $X$, $Z$) are calculated according to the Gaussian distribution:
\begin{eqnarray}
\sigma^2(y_i) = \sum_j (\frac{\partial y_i }{\partial x_j })^2 \sigma^2(x_j)
\end{eqnarray}%
where $(x_1,x_2,x_3,x_4,x_5,x_6)=(lgT_A,lgT_B,R_A,R_B,M_A,M_B)$ and $(y_1,y_2,y_3)=(C_X,X,Z)$. The values of $\sigma(x_j)$ are taken from Table 2. Although the observations of $lgT$, $R$ and $M$ are very accurate, $C_X$ shows a relatively large $\sigma$. The standard errors of the parameters ($C_X$, $X$, $Z$) are mainly due to $\sigma(lgT_B)$, because the contributions of $j=2$ terms in Eq.(8) are $(64\%,44\%,85\%)$ for $i=(1,2,3)$.

\begin{figure}
\includegraphics[scale=0.75]{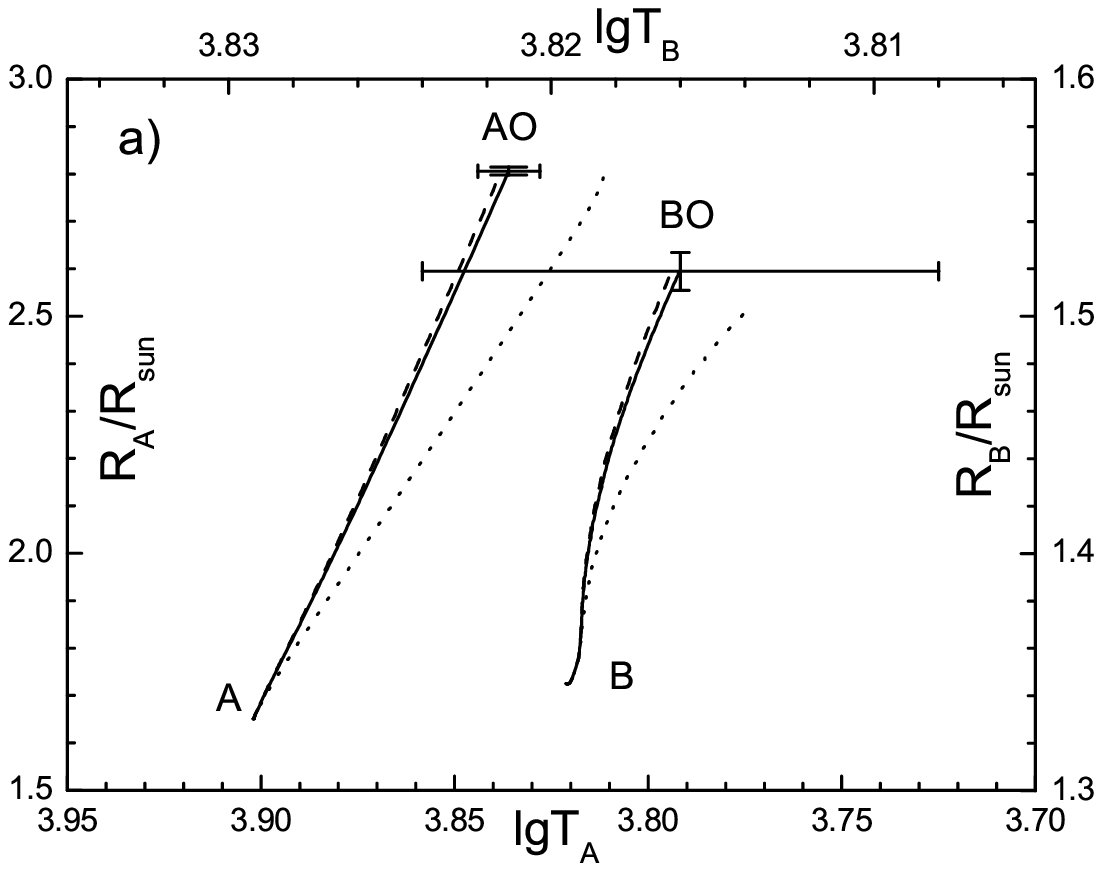}
\includegraphics[scale=0.75]{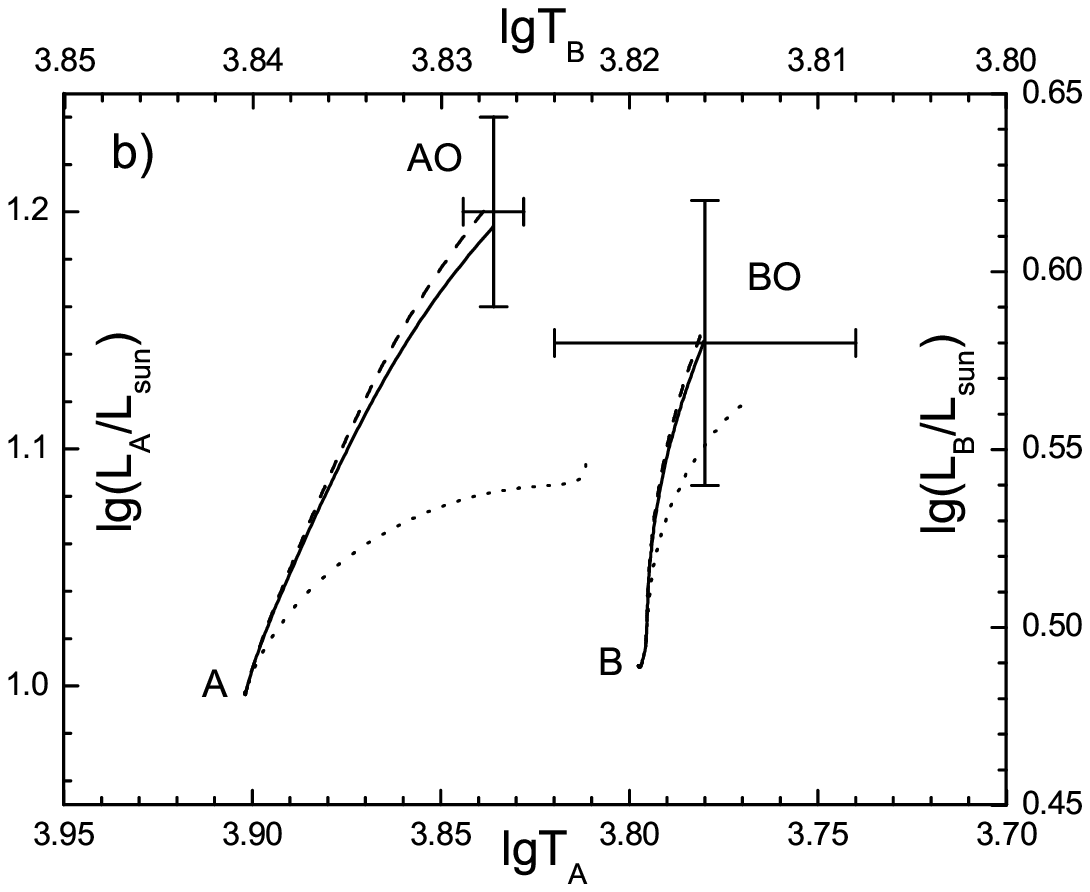}
\caption{The evolutionary tracks of two components of HY Vir with different $C_X$ in the diagram of (a) radius \& (b) luminosity vs. effective temperature. Note that the temperature, luminosity and radius scales are different for the two components of HY Vir. Solid lines correspond to $C_X=0.8\times10^{-10}$, the calibrated model. Dashed lines correspond to $C_X=1.0\times10^{-10}$, the value for the solar case. Dotted lines correspond to the stellar models without overshooting mixing. The $1\sigma$ error bars denoted as 'AO' and 'BO' are based on Table 2. }
\label{Fig.1}
\end{figure}

Figure 1 shows the evolutional tracks of stellar models for both the primary and the secondary, with $X=0.67$, $Z=0.031$ and different $C_X$. Three cases are shown: $C_X=0$ (no overshooting mixing), $C_X=0.8\times10^{-10}$ (the result of calibration) and $C_X=1.0\times10^{-10}$ (suitable for downward overshooting region of the convective envelope in low-mass stars \citep{zha12b,zha12}). It is found that the two cases including the overshooting mixing reproduce the observational effective temperatures and radii in $1\sigma$. Larger $C_X$ results in higher effective temperature and higher luminosity at the termination of the evolution tracks. This is because the overshooting mixing refreshes the hydrogen content of the nuclear burning core and boosts the nuclear reaction in the core.

\section{Properties of the stellar models}

\subsection{The profile of chemical abundance}

\begin{figure}
\includegraphics[scale=0.75]{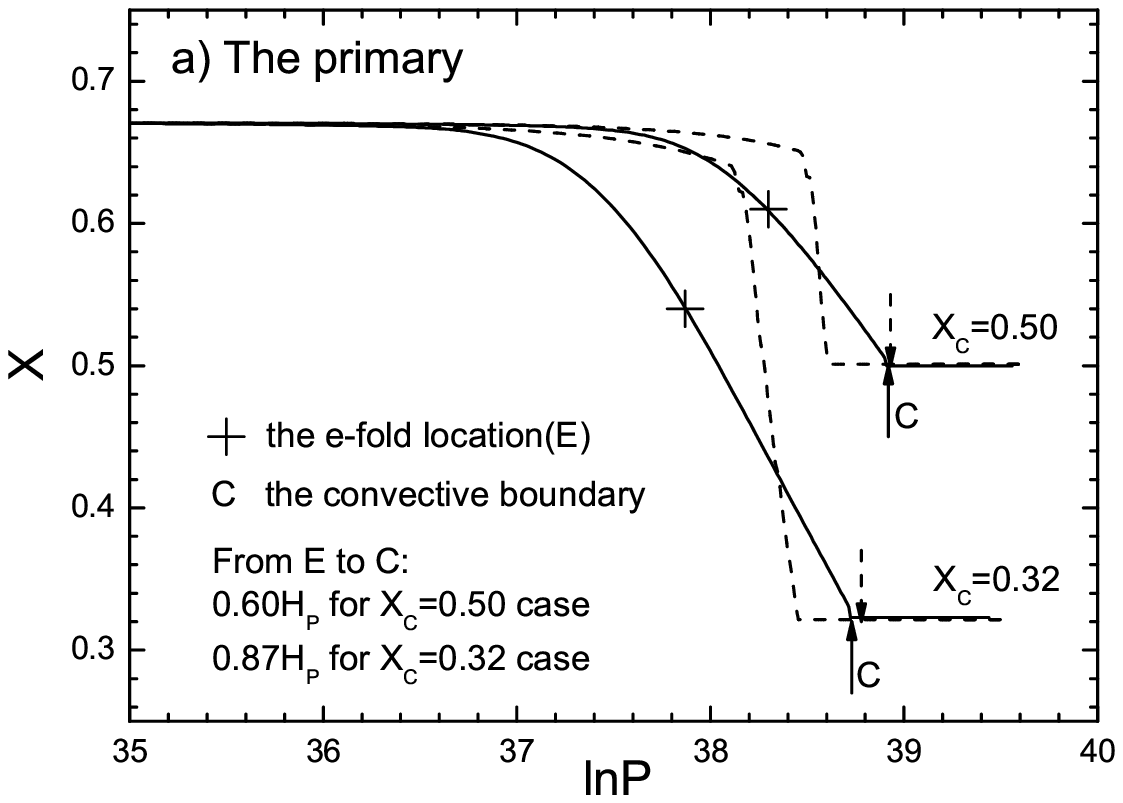}
\includegraphics[scale=0.75]{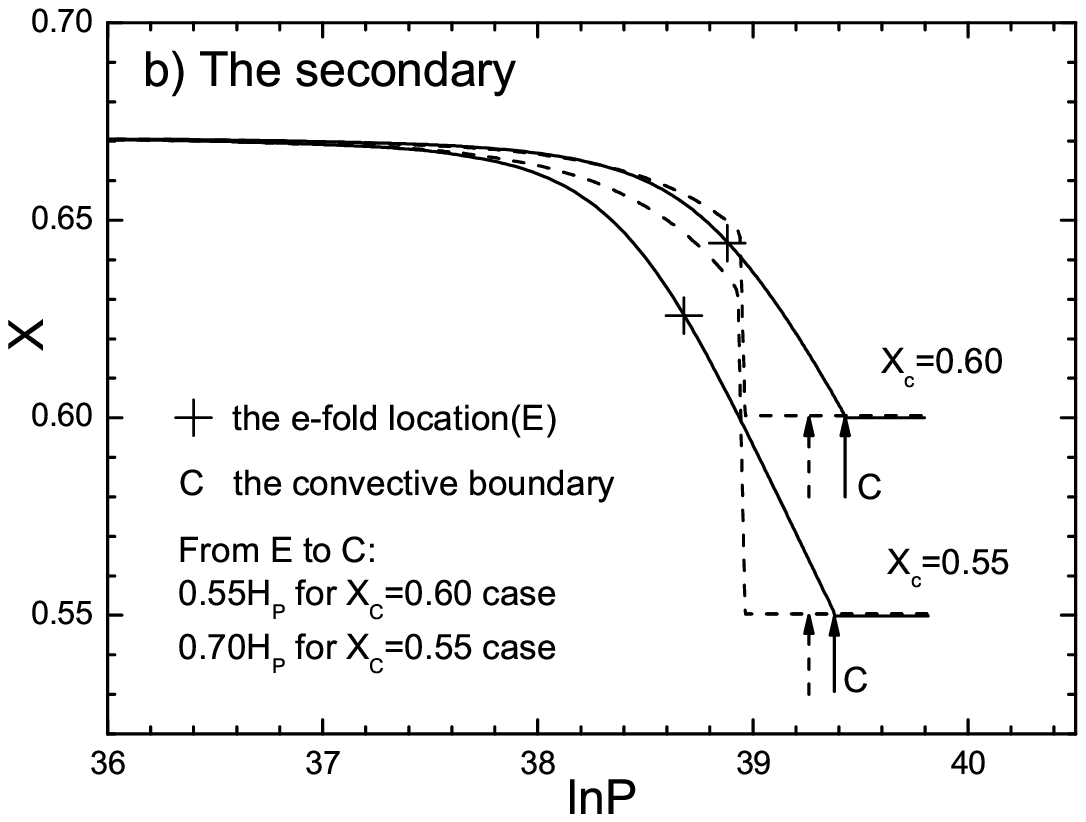}
\caption{The profiles of hydrogen abundance in the stellar interior of (a) the primary stellar models with $X_C = 0.32$ and $X_C = 0.50$, and (b) the secondary stellar models with $X_C = 0.55$ and $X_C = 0.60$. The e-fold location of the chemical profile is defined as the location where $X=X_S+(X_C-X_S)/e$ in which $X_S$ is the hydrogen abundance at the stellar surface. The solid lines correspond to the calibrated stellar models, and the dashed lines correspond to stellar models with $0.3H_P$ fully mixing in overshooting region for comparison. The arrows indicate the convective boundary. }
\label{Fig.2}
\end{figure}

The main effect of the diffusive mixing of the convective overshooting is the modification on the profile of hydrogen abundance.
Figure 2 shows hydrogen abundance profiles of the calibrated stellar models. Stellar models with $X_C=0.50,0.32$ for the primary and stellar models with $X_C=0.60,0.55$ for the secondary are shown, where $X_C$ is the hydrogen abundance of the convective core. The hydrogen abundance profiles of the stellar models with complete mixing in $0.3H_P$ overshooting region are also shown for comparison. The standard convective boundaries are denoted as the arrows in the figure. In the stellar models with complete mixing, the profile is discontinuous at the convective boundary when the convective core expands (e.g., Fig.2(b)) or with high gradient when the convective core slowly contracts (e.g., Fig.2(a)). In the stellar models with the diffusive mixing, $X$ continuously changes from the convective boundary to the outer envelope.

The most important property is the length of the region, in which chemical abundance changes due to the overshooting mixing, since the length indicates the strength of the refreshment of hydrogen in the nuclear burning core. Because the profile of hydrogen abundance continuously changes, it is helpful to discuss by using the concept, i.e., the 'e-folding' length. In the overshooting region, the initial hydrogen abundance can be assumed as the surface hydrogen abundance $X_S$, and the equilibrium abundance is $X_C$. The e-folding length of the chemical abundance changing region is the distance from the convective boundary to the e-folding location defined as $X=X_S+(X_C-X_S)/e$. Figure 2 shows the e-folding length of corresponding stellar models of two components. It is found that the e-folding length increases during the stellar evolution. In order to understand the relation between the e-folding length and the age of stellar model, I estimate the e-folding length. In the concerned evolutionary stage of the components of HY Vir, the stars are in main sequence phase, and the stellar structures are relatively stable (i.e., no rapid core contraction, expansion, ignition of new elements etc.). The convective boundaries show only a little variations. Therefore, one can approximately study the diffusion in a 'quiescent' scene. The time required by the matter diffusing from the convective boundary to the e-fold location should be comparable with the age of the stellar model:
\begin{eqnarray}
t \sim \tau \approx \frac{l_E^2}{4D_{OV}}
\end{eqnarray}%
where $l_E$ is the e-folding length. The diffusion time $\tau \approx l_E^2/(4D_{OV})$ is based on the fundamental solution (i.e., the Green's function) of the basic diffusion equation. Defining $l_E=\alpha_E H_P$, and according to the asymptotical solution of the TCM \citep{zha12b}, one finds that the diffusion coefficient at the e-folding location is as follow:
\begin{eqnarray}
D_{OV}=C_X H_P \sqrt{k_C} exp(-\theta \alpha_E/2)
\end{eqnarray}%
where $k_C$ is the turbulent kinetic energy at the boundary of the convective core, $\theta=dlnk/dlnP$ is the exponential decreasing index of the turbulent kinetic energy in the overshooting region. According to the asymptotical solution, the parameters of the TCM adopted in this paper leads to $\theta=4.8$. Therefore, Eq.(9) can be rewritten as:
\begin{eqnarray}
\alpha_E ^2 exp(\frac{\theta \alpha_E }{2}) \sim 12.6 \frac{ C_{X,-10} W_{C,4} t_G}{H_{P,10}}
\end{eqnarray}%
where $C_{X,-10} = C_X / 10^{-10} $, $W_{C,4} = \sqrt{k_C}/(10^4 cm s^{-1})$, $t_G=t/(1Gy)$ and $H_{P,10}=H_P/(10^{10}cm)$.
Equation (11) indicates that the e-folding length increases during the stellar evolution, which has been found in Fig.2.

\begin{figure}
\includegraphics[scale=0.75]{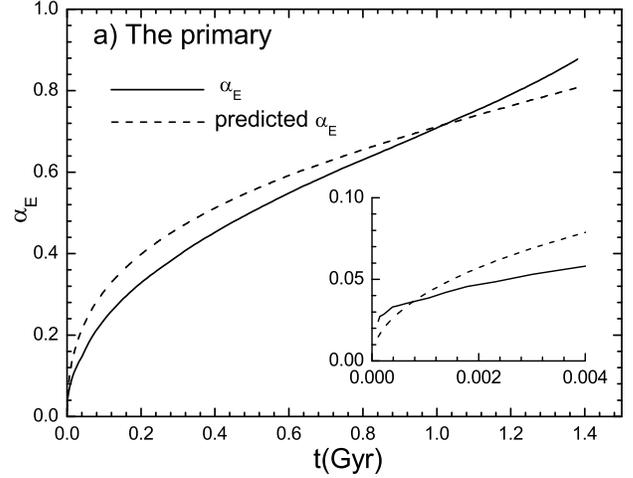}
\includegraphics[scale=0.75]{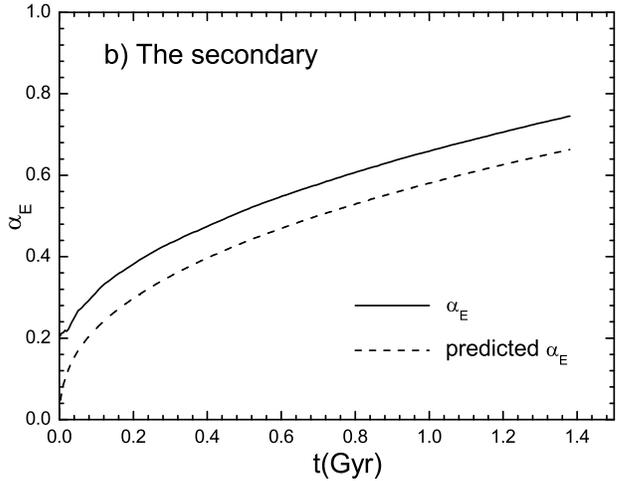}
\caption{The e-fold length parameter $\alpha_E = l_E / H_P$ of the chemical abundance changing region during the stellar evolution. The solid lines are corresponding to the stellar model, and the dashed lines are corresponding to the prediction by Eq.(11). }
\label{Fig.3}
\end{figure}

Figure 3 shows the comparison between the e-folding length predicted by Eq.(11) and the numerical results of stellar models. It is found that the predicted e-folding length is consistent with the numerical results. The predicted e-folding length is zero when $t \rightarrow 0$. However, the e-folding length in stellar models is not zero when $t \rightarrow 0$. This is because the nuclear burning outside the convective core results in a chemical abundance changing region. Equation (11) takes only the overshooting mixing into account, and ignores the nuclear burning. It should be emphasized again that Eq.(11) holds only if the stellar structure is relatively stable. It is invalid for stellar models after the end of the main sequence phase, during which the convective core starts to contract rapidly.

\subsection{The temperature gradient in the overshooting region}

In this paper, the stellar structure equations are solved together with the TCM. In the convection zone, the MLT, which is used in the standard stellar evolutionary calculations, is replaced by the TCM. Comparing with the MLT, the localized TCM shows similar results of temperature gradient in the convection zone. However, the TCM results in non-zero turbulent heat flux $\overline{u_{r}'T'}$ in the overshooting region. The temperature gradient in the overshooting region is modified according to Eq.(4).

\begin{figure}
\includegraphics[scale=0.75]{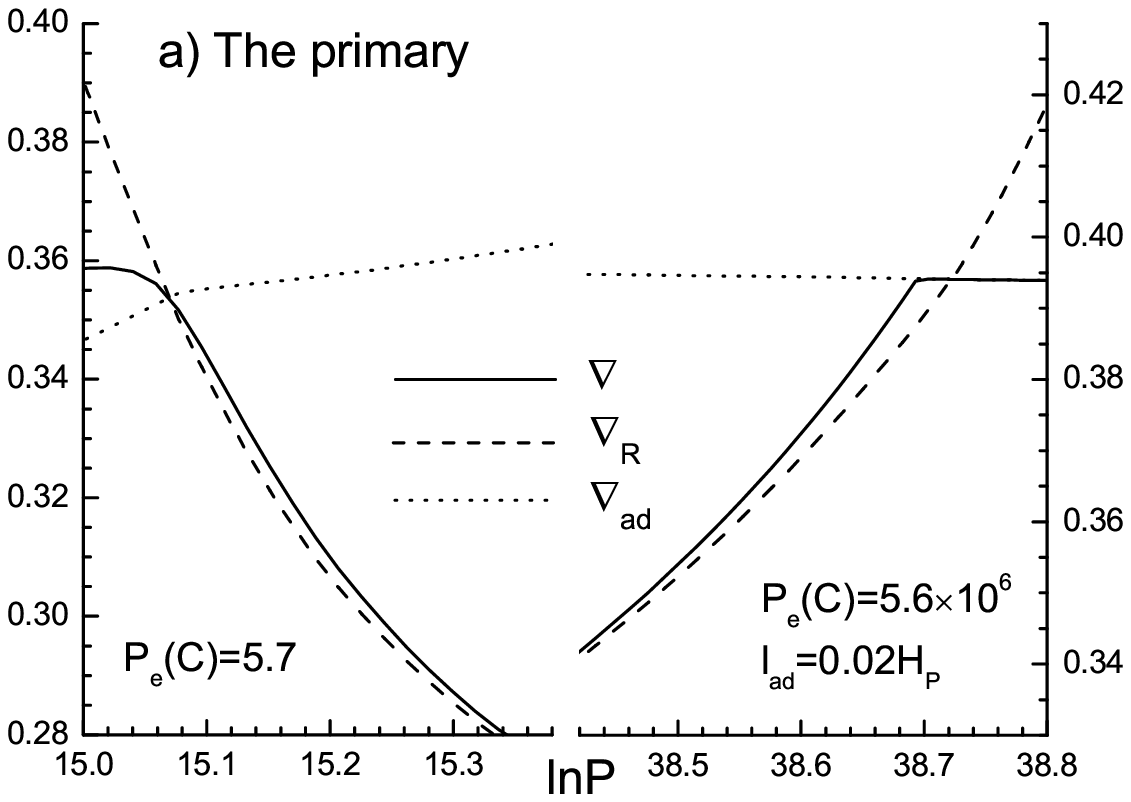}
\includegraphics[scale=0.75]{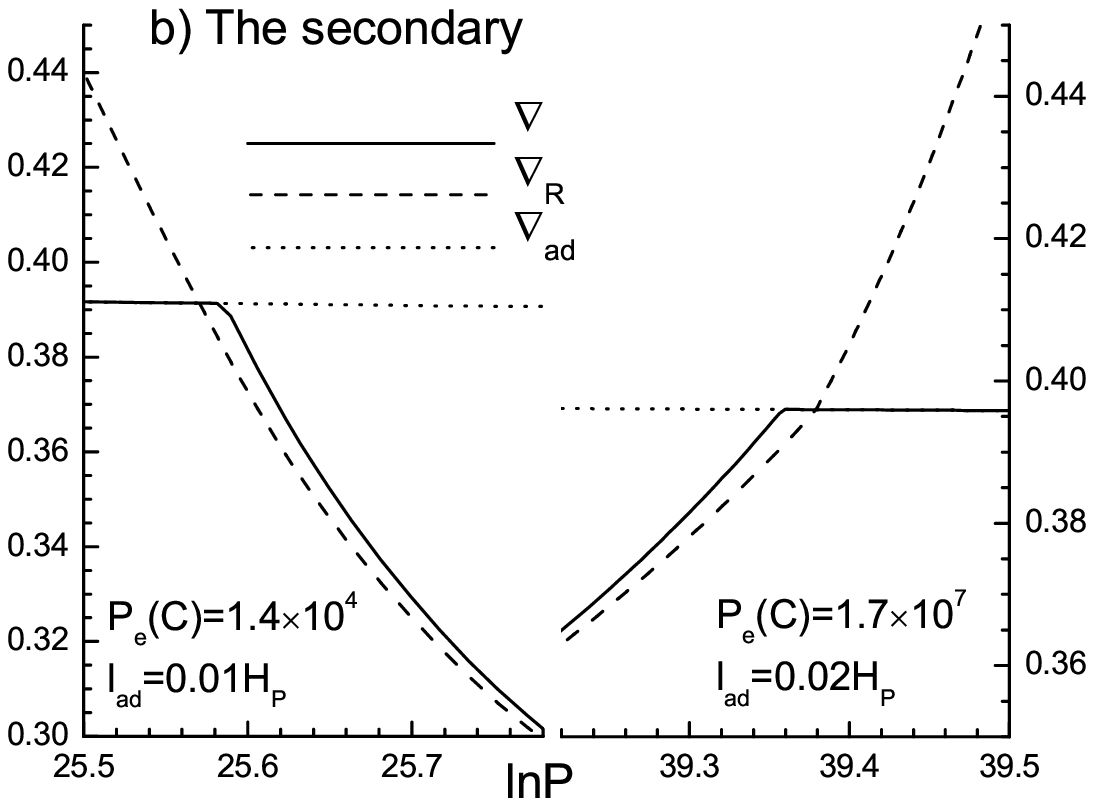}
\caption{The temperature gradient near the convective boundaries. The stellar models are (a) the primary stellar model with $X_C=0.32$ and (b) the secondary stellar model with $X_C=0.55$. The left part of each sub-figure is for the base of the convective envelope, and the right part is for the boundary of the convective core. $P_e(C)$ is the P\'{e}clet number at the convective boundary. $l_{ad}$ is the length of the adiabatic overshooting region.  }
\label{Fig.4}
\end{figure}

Figure 4 shows the temperature gradient near the convective boundaries. It can be found in Fig.4 that $\nabla > \nabla_R$ in the overshooting region. This is a common property of non-local turbulent convection theories \citep{xio85,xio89,can97,xio01,den06,zha09}. These theories show $\overline{u_{r}'T'}<0$ in the overshooting region because the buoyancy prevents turbulent motions, thus $\nabla > \nabla_R$ according to Eq.(4).

It is found in Fig.4 that, in the case of high P\'{e}clet number at the convective boundary, i.e., $P_e(C) \gg 1$, there is a small adiabatic overshooting region adjoining the convective boundary. This has been pointed out by \citet{zha12b} in their theoretical analysis that, in the overshooting region with $P_e \gg 1$, $C_{e1}=0$ (adopted in this paper) results in an adiabatic overshooting region. The length of the adiabatic overshooting region $l_{ad}$ can be estimated by using the parameters of the TCM \citep{zha12b}. The parameters adopted in this paper shows $l_{ad} \approx 0.01H_P$. The diffusion of the turbulent temperature fluctuation $V=\overline{T'T'}$ is absent because of some problems of numerical calculations. If it is taken into account, there should be no adiabatic overshooting region. In most part of the overshooting region even with $P_e(C) \gg 1$, the temperature gradient is close to the radiative one. This may be for the reason that, in the overshooting region with $P_e(C) \gg 1$, the correlativity of turbulent velocity and temperature is almost equal to zero \citep{xio85,xio01,den06,zha12b}, thus turbulent motions can hardly transport heat.

Although the TCM modifies the temperature gradient in the overshooting region, the modification is small. Zhang \& Li's (2012a) results have indicated that, comparing with the overshooting mixing, the temperature gradient modification based on the adopted TCM parameters is ignorable on affecting the stellar structure and evolution.

\subsection{The turbulent variables in the overshooting region}

The turbulent variables involved in the TCM are as follows: $\overline{u_{r}'u_{r}'}$ is the radial turbulent kinetic energy, $k$ total turbulent kinetic energy, $\overline{u_{r}'T'}$ the turbulent heat flux, $\overline{T'T'}$ the turbulent temperature fluctuation.

\begin{figure}
\includegraphics[scale=0.75]{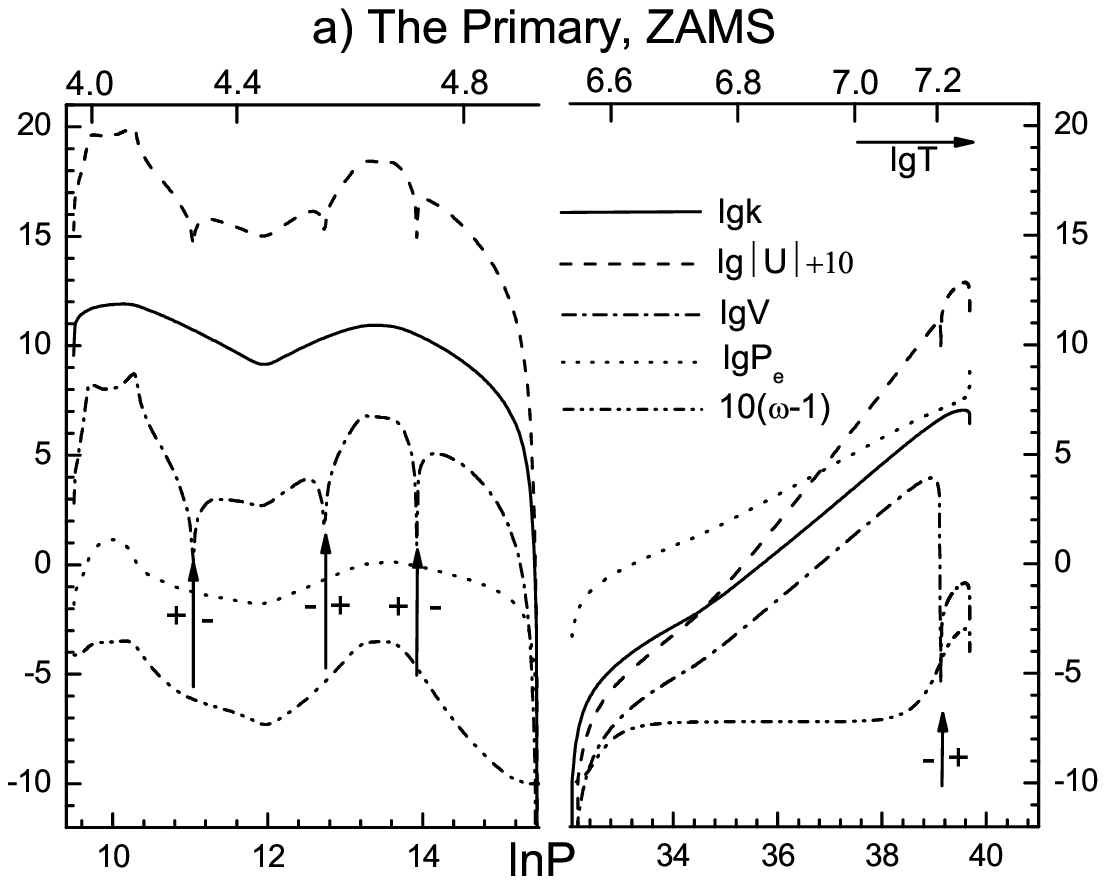}
\includegraphics[scale=0.75]{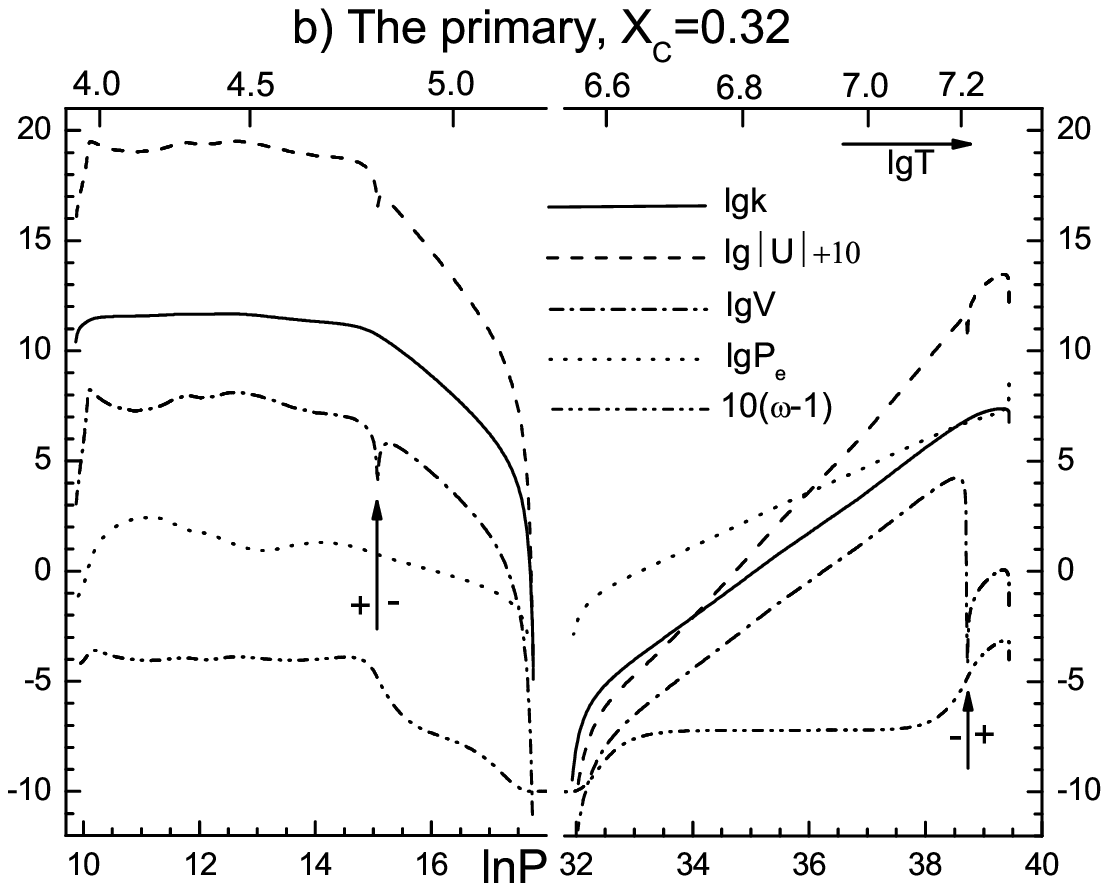}
\caption{The turbulent variables of the stellar models of the primary. $k$ is turbulent kinetic energy, $U=\overline{u_{r}'T'}$ the turbulent heat flux, $V=\overline{T'T'}$ the turbulent temperature fluctuation, $P_e$ the P\'{e}clet number, $\omega = \overline{u_{r}'u_{r}'} / (2k)$ the anisotropic degree. The arrows indicate the convective boundary. '+' and '-' around the arrows indicate the convection zone and overshooting region, respectively. }
\label{Fig.5}
\end{figure}

\begin{figure}
\includegraphics[scale=0.75]{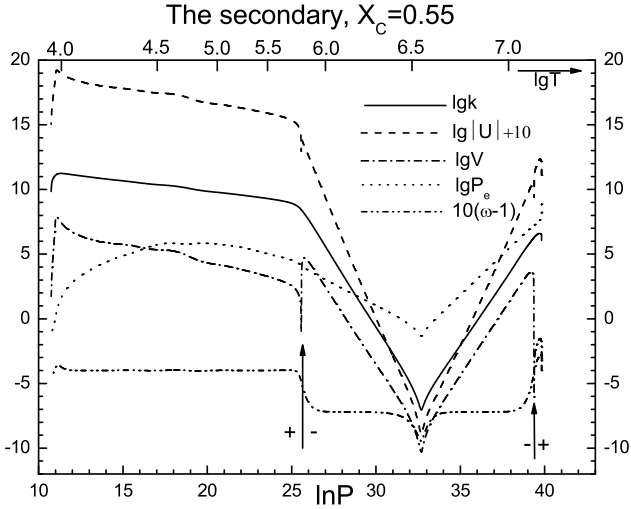}
\caption{Similar to Fig.5, but for the stellar models of the secondary. }
\label{Fig.6}
\end{figure}

The overshooting region can be classified into two kinds: the high $P_e$ overshooting region with $P_e \gg 1$ and the low $P_e$ one. The core overshooting region is always the high $P_e$ one because of the short free path of photons in the stellar interior. The type of the downward overshooting region of the convective envelope depends on the temperature at the convective boundary $T_{BCE}$. Commonly, it is the high $P_e$ overshooting region for $lgT_{BCE}>5.5$ or the low $P_e$ one for $lgT_{BCE}<5.5$. The properties of the two kinds of overshooting region have been found to be different with each other \citep{xio92,xio01,zha12b}. The properties of the high $P_e$ overshooting region in the framework of Li \& Yang's (2007) TCM have been studied by \citet{zha12b}. However, the property of the low $P_e$ overshooting region is not as clear as the high $P_e$ one.

Figures 5-6 show the turbulent variables in the stellar interior. For the primary, the ZAMS model and the model with $X_C=0.32$ are shown. For the secondary, only the stellar model with $X_C=0.55$ is shown, because the structure of the turbulent variables shows only a little variations during the concerned evolutionary range. At all convective boundaries, $\overline{u_{r}'T'}$ and $\overline{T'T'}$ are equal to zero because their diffusions are ignored in the calculations.

As shown in Figs.5-6, $\overline{T'T'}$ in the core overshooting region is larger than in the core convection zone. This is coincident with \citet{xio85}. The turbulent temperature fluctuation $\overline{T'T'}$ represents the difference of the temperature between the turbulent flows and the environment, i.e., $\triangle T$. There is always $P_e \gg 1$ in the core convection zone. In the $P_e \gg 1$ convection zone, the turbulent heat transport is very efficient, leading to $\nabla \approx \nabla_{ad}$. Therefore, the temperature of the turbulent flows is almost equal to the temperature of the environment, i.e., $\triangle T\approx0$, thus $\overline{T'T'}\approx0$ in the $P_e \gg 1$ convection zone. In the overshooting region, however, $\nabla$ is close to $\nabla_R$ so that $\triangle T$ is significant.

\begin{table}
\begin{center}
\caption{The properties of the turbulent variables in the high $P_e$ overshooting region: numerical results vs. theoretical predictions. }
\begin{tabular}{crrrrr}
  \hline
     & C Fig.5(a) & C Fig.5(b) & C Fig.6 & E Fig.6 & Asy. \\
  \hline
  $\sqrt{k_C}$   & $2060$& $2760$& $1410$& $16500$&-  \\
  $\sqrt{k_C}$(TMOD)   &$2120$&$2510$&$1630$&$15200$&-  \\
  \hline
  $dlnk/dlnP$   & $4.5$ & $4.4$& $4.6$& $-4.8$& $\pm4.8$ \\
  $dln(-\overline{u_{r}'T'})/dlnP$      &$6.8$&$7.1$&$6.7$&$-7.2$& $\pm7.2$  \\
  $dln(\overline{T'T'})/dlnP$   &$4.6$&$4.5$&$4.5$&$-4.6$& $\pm4.8$   \\
  $\omega = \overline{u_{r}'u_{r}'} / (2k)$   & $0.28$ & $0.28$& $0.28$& $0.28$& $0.28$  \\
  \hline
\end{tabular}
\tablecomments{$k_C$ is the turbulent kinetic energy at the convective boundary. The 'Asy.' column show the results predicted by the asymptotical analysis\citep{zha12b}. '$\sqrt{k_C}$ (TMOD)' is the value of $\sqrt{k_C}$ predicted by the method called 'the maximum of diffusion'\citep{zha12b}. The 'C Fig.5(a)' column means the core overshooting in Fig.5(a), and the 'E Fig.6' column means the envelope overshooting in Fig.6. }
\end{center}
\end{table}

In the high $P_e$ overshooting region (in the right parts of Fig.5(a-b), and in Fig.6), the turbulent variables $k$, $\overline{u_{r}'u_{r}'}$, $\overline{T'T'}$ decrease exponentially, and the anisotropic degree $\omega = \overline{u_{r}'u_{r}'} / (2k)$ almost doesn't change. Those are consistent with the asymptotical analysis of the TCM \citep{zha12b} and Xiong's non-local turbulent model (see, e.g., \citet{xio85,xio89,xio01,den08}). Table 3 compares the properties of the turbulent variables in the high $P_e$ overshooting region with the results of the theoretical analysis \citep{zha12b}. It is found that the numerical results are in agreement with the theoretical analysis. The high $P_e$ overshooting region extends to the location of $P_e \sim 1$ and then becomes the low $P_e$ one.

In the low $P_e$ overshooting region, the thermal dissipation dominates, thus the turbulent variables $k$, $\overline{u_{r}'u_{r}'}$, $\overline{T'T'}$ decrease super-exponentially and $\omega$ decreases to zero according to the properties of the TCM equations \citep{zha12b}. Those properties can be found in Fig.5. The overshooting is finally cut-off in the region $P_e \ll 1$. According to the definition of $P_e$, $P_e \propto \sqrt{k}$, thus $P_e$ exponentially decreases in the overshooting region with the index $\theta /2$. As a consequence, the distance (in $H_P$) from the convective boundary to the cut-off location depends on $P_e(C)$. Larger $P_e(C)$ leads to longer distance to the cut-off location. This can be found in Figs.(4-6) by comparing $P_e(C)$ and the the distance from the convective boundary to the cut-off location of each overshooting region. 

\begin{figure}
\includegraphics[scale=0.75]{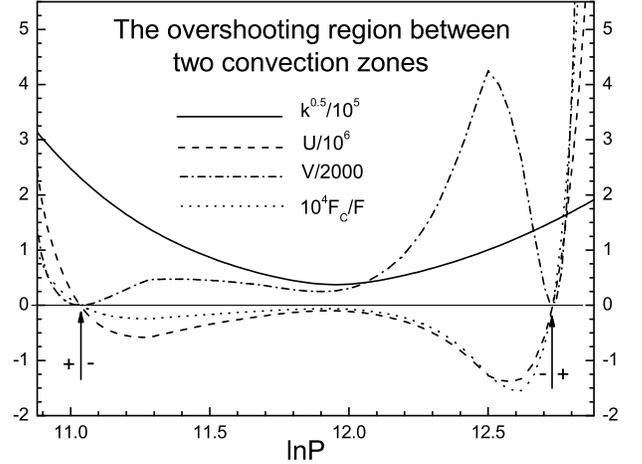}
\caption{The turbulent variables in the overshooting region which lays between two convection zones. The dotted line is the ratio of the convective flux (see text) to the total flux $F_C/F = 1-\nabla/\nabla_R$. The stellar model is the ZAMS of the primary. Symbols are the same as Fig.5. }
\label{Fig.6}
\end{figure}

There is a thin overshooting region laying between two convective shells, which are caused by H and HeII ionization respectively, in the ZAMS stellar model of the primary as shown in Fig.5a. The convective shells merge together during the stellar evolution, thus this overshooting region disappears in the $X_C=0.32$ stellar model. The details of the turbulent variables in the thin overshooting region are shown in Fig.7. The turbulent velocity $\sqrt{k}\sim10^5 cm/s$ in the region is larger than in the convective core, so that two convective envelopes can be considered to be dynamically connected. However, the minimum of the ratio of the convective flux to the total flux is about $|F_C/F| = |1-\nabla/\nabla_R| \sim 10^{-5}$, where $F_C=\rho c_P \overline{u_{r}'T'}$ is the convective flux. Moreover, $P_e<1$ in the two convective shells and the overshooting region. These conditions prevent the turbulent heat transport between two convective shells. The convective envelopes can therefore be considered to be thermally disconnected.

\section{Conclusion and discussion}

In this work, the turbulent convection model (TCM) is involved in the stellar structure and evolution in order to study the convective core overshooting in the stellar models of eclipsing binary star HY Vir. Using new accurate observations, I calibrate the stellar effective temperatures and radii of two components of HY Vir, and then obtain the required overshooting mixing parameter $C_X=(0.8\pm0.5)  \times10^{-10}$. The turbulent parameters (i.e., the TCM parameters and the overshooting mixing parameter $C_X$), which is proper for the downward overshooting region of the convective envelope in low-mass stars, can reproduce the observational radii and effective temperatures of the two components of HY Vir. It indicates that the diffusive overshooting mixing based on the turbulent velocity described by the TCM can also be applied to the core overshooting. This result encourages us to apply the TCM to the convective overshooting.

The diffusive overshooting mixing causes a continuous profile of the hydrogen abundance. The e-folding length of the region, in which the chemical abundance changes due to the overshooting mixing, can be estimated by using Eq.(11). It is found that the e-folding length increases during the stellar evolution.

The overshooting region can be classified into two kinds: the high $P_e$ overshooting region in which $P_e \gg\ 1$ and the low $P_e$ one. The properties of the high $P_e$ overshooting region are in agreement with the theoretical analysis \citep{zha12b}. The turbulent variables (turbulent kinetic energy $k$, turbulent heat flux $\overline{u_{r}'T'}$ and turbulent temperature fluctuation $\overline{T'T'}$) decrease exponentially in the high $P_e$ overshooting region. The properties of the low $P_e$ overshooting region are very different from the high $P_e$ one. In the low $P_e$ overshooting region, the turbulent variables decrease super-exponentially and are finally cut-off.

The diffusions of $\overline{u_{r}'T'}$ and $\overline{T'T'}$ are not taken into account in this paper because of some problems of numerical calculations. However, $C_X$ of the calibrated stellar models for HY Vir should not change significantly if those diffusions are present. Those diffusions may affect the calibrated results mainly on two ways: (i) modifying the turbulent heat flux and (ii) affecting the core overshooting mixing which plays an important role on the stellar evolution. Although those diffusions can redistribute $\overline{u_{r}'T'}$ in the stellar interior, they should't significantly change the integral value of $\overline{u_{r}'T'}$. The effective temperature of stellar models should be insensitive to those diffusions. The diffusion of $\overline{u_{r}'T'}$ is ignorable in the region of $P_e \gg 1$ \citep{zha12b}, thus it doesn't affect the properties of the core convection and overshooting. The diffusion of $\overline{T'T'}$ can affects the exponential index of the turbulent kinetic energy $\theta(=dlnk/dlnP)$. The solar value for the diffusion of $\overline{T'T'}$ is small that $C_{e1}=0.02$ \citep{zha12a}. It should modify the turbulent properties in the core overshooting region slightly. $C_X$ of the calibrated model should change a little if the diffusion of $\overline{T'T'}$ is present. However, since the relative standard error of $C_X$ is about 60\%, the solar value $C_X=10^{-10}$ should also reproduce the required radius and effective temperature of HY Vir in $1\sigma$.

In other treatments of overshooting with incomplete mixing mentioned in Section 2.1, there is no information on how the temperature gradient profile is modified near the convective boundary. The helioseismic investigation by \citet{chr11} can't judge which of them is suitable for the solar case. However, the mixing can affect the sound speed profile. It is well known that, in the standard solar model, there is a bump of sound speed difference below the solar convection zone. In order to eliminate the bump, the proper diffusion coefficient at the base of the solar convection zone is on the magnitude order of $10^2 \sim 10^3$ (see, e.g., \citet{chr07,zha12a}). The mixing should be effective in about $0.1R$, which is the width of the bump, below the base of convection zone. The helioseismic inversion requires the diffusion coefficient to satisfy those conditions in the solar case.

\acknowledgments

Many thanks to the referee for providing productive and valuable comments. This work is co-sponsored by the National Natural Science Foundation of China through grant No.10973035, Science Foundation of Yunnan Observatory No.Y0ZX011009 and No.Y1ZX011007, and Chinese Academy of Sciences under grant no. KJCX2-YW-T24.

\end{document}